\documentclass[12pt,titlepage]{article}
\usepackage[authoryear]{natbib}
\usepackage{amsfonts}
\usepackage{amsmath}
\usepackage{amssymb}
\usepackage{graphicx}
\usepackage{rotating}
\usepackage{marginnote}
\usepackage{color}
\usepackage{fullpage}

\newcommand{\E}{\mathbb{E}}
\renewcommand{\P}{\mathbb{P}}

\begin{document} 

%% shorter header: A partial draft of hitchhiking
\title{Patterns of neutral diversity under general models of selective sweeps}
%The effect of recurrent partial sweeps on patterns of neutral diversity.}
\author{Graham Coop$^{1}$ and Peter Ralph$^{1}$\\ 
\small $^1$ Department of Evolution and Ecology \& Center for Population Biology,\\
\small University of California, Davis.\\
\small To whom correspondence should be addressed: \texttt{gmcoop@ucdavis.edu}\\
}
\date{}
\maketitle

\section*{Abstract}
Two major sources of stochasticity in the dynamics of neutral alleles result from resampling of finite populations (genetic drift) and the random genetic background of nearby selected alleles on which the neutral alleles are found (linked selection). There is now good evidence that linked selection plays an important role in shaping polymorphism levels in a number of species. 
One of the best investigated models of linked selection is the recurrent full sweep model, in which newly arisen selected alleles fix rapidly. 
However, the bulk of selected alleles that sweep into the population may not be destined for rapid fixation. 
Here we develop a general model of recurrent selective sweeps in a coalescent framework,
one that generalizes the recurrent full sweep model to the case where selected alleles do not sweep to fixation. 
We show that in a large population, only the initial rapid increase of a selected allele affects the genealogy at partially linked sites, 
which under fairly general assumptions are unaffected by the subsequent fate of the selected allele. 
We also apply the theory to a simple model to investigate the impact of recurrent partial sweeps on levels of neutral diversity,
and find that for a given reduction in diversity, 
the impact of recurrent partial sweeps on the frequency spectrum at neutral sites is determined primarily by the frequencies achieved by the selected alleles.
Consequently, recurrent sweeps of selected alleles to low frequencies can have a profound effect on levels of diversity but can leave the frequency spectrum relatively unperturbed. 
In fact, the limiting coalescent model under a high rate of sweeps to low frequency is identical to the standard neutral model. 
The general model of selective sweeps we describe goes some way towards providing a more flexible framework to describe genomic patterns of diversity than is currently available.
 
\section{Introduction}

The high levels of genetic variation within natural populations have long fascinated population geneticists. One school of thought holds that a substantial proportion of this molecular polymorphism is neutral or very weakly deleterious \citep{KimuraOhta:71,Ohta:73,Kimura:book}. For neutral polymorphism, the level of genetic diversity results from a balance between the introduction of alleles through mutation and their stochastic loss \citep{KimuraCrow:1964,Kimura:1969,Ewens:1972}. Under the neutral theory of molecular evolution this stochasticity is thought to result mostly from genetic drift \citep{Kimura:book}, the random resampling that occurs in finite populations, an effect that is exaggerated by fluctuating population size and large variation in reproductive success among individuals \citep[see][for a recent review]{Charlesworth:09}. However, selection at linked sites may provide a major source of stochasticity as the dynamics of a neutral allele can be strongly influenced by the random genetic background on which selected alleles arise \citep{MaynardSmith:74,Kaplan:89,Charlesworth:95,Hudson:95}. 

In many species examined to date, levels of diversity are substantially lower in regions of low recombination, as found in multiple species of \emph{Drosophila} \citep{Aguade:89,Begun:92,Berry:91,Shapiro:07,Begun:07}, \emph{Caenorhabditis} \citep{Cutter:03,Cutter:10}, humans \citep{Hellmann:08,Cai:09} and \emph{Saccharomyces cerevisiae} \citep{Cutter:11}; but not in all species, e.g.\ \emph{Arabidopsis} \citep[][]{Nordborg:05,Wright:06}. Moreover, levels of diversity are also lower in regions that {\it a priori} are expected to have a higher rate of functional mutations, e.g.\ near genes and conserved elements \citep{Mcvicker:09,Cai:09,Hernandez:11}. 
Since the rate of neutral genetic drift is independent of recombination rate, 
% These observations are inconsistent with genetic drift being the sole source of stochasticity in allele frequencies because the rate of drift should be independent of the recombination rate. 
this positive correlation between recombination rates and diversity offers good evidence that linked selection plays a substantial role in the fate of alleles, especially in low recombination regions. What is still far from clear is how different forms of linked selection contribute to this reduction, and whether linked selection can explain the narrow observed range of genetic diversity across species with vastly different (census) population sizes \citep{Lewontin:74,MaynardSmith:74}. 

Models of the effect of linked selection have often been divided between those that propose the source of this linked selection to be either the purging of deleterious variation (background selection) or the selective sweep of beneficial alleles (hitchhiking). In this paper we explore the consequences of a generalized model of hitchhiking on patterns on neutral diversity. We first review some of the key results of models of linked selection. Under the background selection model, genetic diversity is continuously lost from natural populations due to the removal of haplotypes that carry deleterious alleles \citep{Charlesworth:95,Hudson:95}. For strongly deleterious alleles, this continuous loss acts primarily to increase the rate of genetic drift at markers closely linked to loci with high deleterious mutation rates \citep{Hudson:95b,Nordborg:96}. Therefore, this background selection model leads to a reduction in genetic diversity but no skew in the frequency spectrum. However, a skew towards rare neutral alleles can result if weakly deleterious mutations are incorporated into the model \citep{Nordborg:96, Gordo:02}.
 
On the other end of the spectrum, \citet{MaynardSmith:74} proposed that local levels of genetic diversity could be reduced by the hitchhiking effect. The hitchhiking effect results from the fact that when an initially rare, beneficial allele sweeps rapidly to fixation it carries with it a linked region of the haplotype on which it arose. The size of genomic region affected by a recent sweep is proportional to the ratio of the strength of selection to the rate of recombination \citep{MaynardSmith:74,Kaplan:89, Stephan:92, Barton:98}, and so the reduction in levels of diversity is determined by the distribution of selection coefficients and the rate of sweeps per unit of the genetic map.  Neutral alleles further away from the selected site may not be pulled all of the way to fixation if recombination occurs during the sweep, which can lead to a transient excess of high-frequency derived alleles an intermediate distance away from the selected site after each sweep \citep{Fay:00,Przeworski:02,Kim:06}. As neutral diversity levels slowly recover through an influx of new mutations after the sweep there is a strong skew towards low frequency derived alleles, a pattern that persists for many generations \citep{Braverman:95, Przeworski:02,Kim:06}.  In a large population, the rate of sweeps could be high enough that hitchhiking dominates genetic drift as the source of stochasticity \citep{MaynardSmith:74,Kaplan:89,Gillespie:00}, an idea which has been termed genetic draft \citep{Gillespie:00}. 

Support for a hitchhiking model over the standard model of background selection  is found in \textit{Drosophila}, where there is a greater skew towards rare alleles at putatively neutral sites in regions of low recombination \citep{Andolfatto:01,Shapiro:07} and regions surrounding amino-acid substitutions have lower levels of diversity \citep{Andolfatto:07, Macpherson:07,Sattath:11}. However, in humans (and other species) there is no strong skew towards rare alleles in low recombination regions \citep{Mcvicker:09,Hernandez:11,Lohmueller:11}, which combined with other evidence \citep{Coop:09, Hernandez:11} suggests that full sweeps may have been rare, and that background selection may be the main mode of linked selection, in humans and a number of other species.

Although the recurrent full sweep model has been the subject of considerable theoretical investigation, it may actually be relatively rare for advantageous alleles to sweep rapidly all the way to fixation. Fluctuating environments \citep[e.g.][]{Gillespie:91, Kopp07,,Kopp09a,Kopp09b} and changing genetic backgrounds may often act to prevent alleles achieving rapid fixation within the population (see \citet{Pritchard:10} for a recent discussion). For example, if multiple mutations affecting the adaptive phenotype segregate during the sweep then it may be that no one of these alleles sweeps to fixation \citep{Pennings:06,Pennings:06b,Chevin:08,Ralph:10}. Multiple alleles spreading rapidly from low frequency can lead to either a set of partial sweeps within the population, or a soft sweep if the alleles are tightly linked.  Furthermore, a similar effect can occur when selection acts on an allele present as standing variation, if the allele is present on multiple haplotypes when it starts to spread \citep{Innan:04,Hermisson:05,Przeworski:05}. The fact that, under these models, no single haplotype goes quickly to fixation acts to reduce the hitchhiking effect, and alters the effect on the frequency spectrum. 

The genome-wide effect of other modes of linked selection on patterns of diversity is relatively unexplored. One model that has been investigated is an infinitesimal model of directional selection, where the aggregated effect of selection over many loci can be a substantial source of stochasticity at linked and even unlinked sites \citep{Robertson:61,Santiago:95,Santiago:98,Barton:00}. Fluctuating selection due to varying environments has also been shown to lead to reduced levels of diversity at linked neutral sites \citep{Gillespie_nonneutral,Gillespie:97,Barton:00} and simulations of specific models of fluctuating selection have shown that the same reduction in diversity can result in a much smaller skew in the frequency spectrum than under the hitchhiking model \citep{Gillespie_nonneutral,Gillespie:97}. However, as yet no coalescent model of the effect of recurrent incomplete sweeps has been developed. 

Here is an outline of how we proceed.
First, we develop a coalescent-based model of patterns of diversity surrounding a selected allele that sweeps into the population but not necessarily to fixation. We concentrate on the case of a very large population and sites that are partially linked to this selected locus. We find that if the initial rise of the selected allele is rapid then the coalescent process is primarily affected by this stage, and relatively insensitive to the subsequent dynamics of the selected allele. Using this intuition, we then develop a coalescent model of recurrent sweeps on patterns of neutral diversity in which selected alleles may only reach intermediate frequency. To test the approximations involved in the model we compare the results at several stages to simulations.  Some of the implications of these results for interpretation of genome-wide diversity patterns are presented in the discussion.   

\section{Results}
\subsection{Coalescent framework and assumptions} \label{trajectory_section}
As first described by \citet{Kaplan:88} and \citet{Hudson:88}, patterns of neutral diversity at a neutral locus linked to a selected locus can be modeled by 
conditioning on the trajectory of the frequency of the selected allele through time, and
treating the two allelic classes as subpopulations within each of which the dynamics are neutral,
with recombination moving lineages between the two \citep[see also][]{Barton:04, Barton:04b}. 
Consider a locus under selection at which a derived allele $D$ and an ancestral allele $A$ segregate,
and let the frequency of $D$ at time $t$ be denoted $X(t)$.  
We will study the coalescent process at a neutral locus partially linked to our selected locus, 
with recombination occurring at rate $r$ per generation between the selected and the neutral locus. 
Each ancestor on a given lineage in the coalescent process carried either the $D$ or the $A$ allele at the selected locus,
which we refer to as the ``type'' of that lineage. %deleted 'with' NICK

Throughout we assume that the diploid population size $N$ is large and constant over time.
For simplicity, we assume that the effective population size is $2N$,
(i.e.\ the neutral coalescence rate of a pair of lineages is $1/(2N)$)
and that no more than two lineages coalesce at once in the absence of a selective sweep.

Suppose at time $t$ that $k_D$ and $k_A$ of our lineages are of the derived and ancestral type respectively.
There are $NX(t)$ individuals carrying the derived allele that could be progenitors of the $k_D$ lineages,
so the instantaneous rates of coalescence of pairs of lineages within the two allelic classes at time $t$ are
\begin{equation}
  {k_D \choose 2} \frac{1}{2NX(t)} \qquad \mbox{and} \qquad {k_A \choose 2} \frac{1}{2N(1-X(t))}, \qquad \mbox{respectively.} \label{allelic_coal_rate}
\end{equation}
The total instantaneous rate of recombination is $(k_D+k_A) r$. If a recombination event occurs on a lineage at time $t$,
it chooses to be of type $D$ with probability $X(t)$, and chooses to be of type $A$ otherwise.

We will leave the dynamics of the selective sweeps that determine $X(t)$ fairly unspecified,
and while stochasticity may play an important role in shaping the trajectories, in examples we usually treat $X(t)$ as nonrandom. 
As we want coalescences caused by a single selective sweep to occur at more or less the same time,
we require that once the selected allele is introduced into the population it increases in frequency rapidly,
and that once the allele frequency leaves the boundary (e.g.\ moves above 1\%),
it does not return (e.g.\ drops below 1\%)
unless it does so on the way to loss (e.g.\ hits 0 before returning to 1\%). 
This condition implies that our model applies to alleles that are at least partially codominant, as fully recessive alleles spend appreciable time, behaving stochastically, at very low frequencies, which can lead to different coalescent dynamics at linked loci \citep{Teshima:06a,Ewing:11}.

%%%%%%%%%%%
\subsection{Relation to previous models}

We describe a simple approximation to the coalescent with recurrent sweeps that is inspired by similar approximations for a model of recurrent full sweeps. The approximation postulates two types of coalescent events -- ``neutral'' events occurring at rate $1/2N$ between any pair of lineages, and additional coalescent events, involving two or more lineages, due to selective sweeps. 
The first class of events can occur at any time, due to random resampling of lineages.
The second class of events, the sweep--induced coalescent events, can involve more than two lineages, 
as we assume that lineages forced to coalesce by a sweep do so instantaneously on the relevant time scale. 
We assume that all such lineages coalesce into a single lineage,
and that the distribution of the number of such lineages is binomial, 
with a success probability that is a function of the trajectory taken by the selected allele and the recombination distance to that allele.
This framework is a natural extension of similar approximations used for full sweeps \citep{Barton:98,Gillespie:00,KimStephan:02, Nielsen:05,Durrett:05}.

Processes with two classes of coalescent events have previously been developed to approximate a recurrent full-sweep model \citep{Kaplan:89,Gillespie:00,Durrett:05}. 
When the transition probabilities can be written in this binomial form, as they also are in the recurrent full sweep models of \citet{Gillespie:00} and \citet{Durrett:05}, 
the model is called a $\Lambda$-coalescent \citep{Pitman:99,Sagitov:99}. 
These also arise in neutral models where individuals have large variance in reproductive success \citep[e.g.][]{SargsyanWakeley:08,MohleSagitov:01}. 
As in other work, we present this model as an approximation not in the sense of asymptotic convergence,
but rather as a simplification, which we show later is close enough to be useful. 
We make a number of simplifying assumptions, and often do not make use of the most accurate analytical forms available,
in an effort to maintain an intuitive form and description of the process obtained.
In particular, \citet{Durrett:04} showed that a coalescent process with simultaneous multiple collisions could provide a better approximation to the coalescent process during a sweep,
a direction we do not pursue \citep[see also][]{Barton:98,Etheridge:06}.

%%%%%%%%%%%%%%%%%%%%%%%%
\subsection{An approximation to the coalescent process during the sweep}

\begin{figure} 
  \includegraphics[angle=90,width=\textwidth]{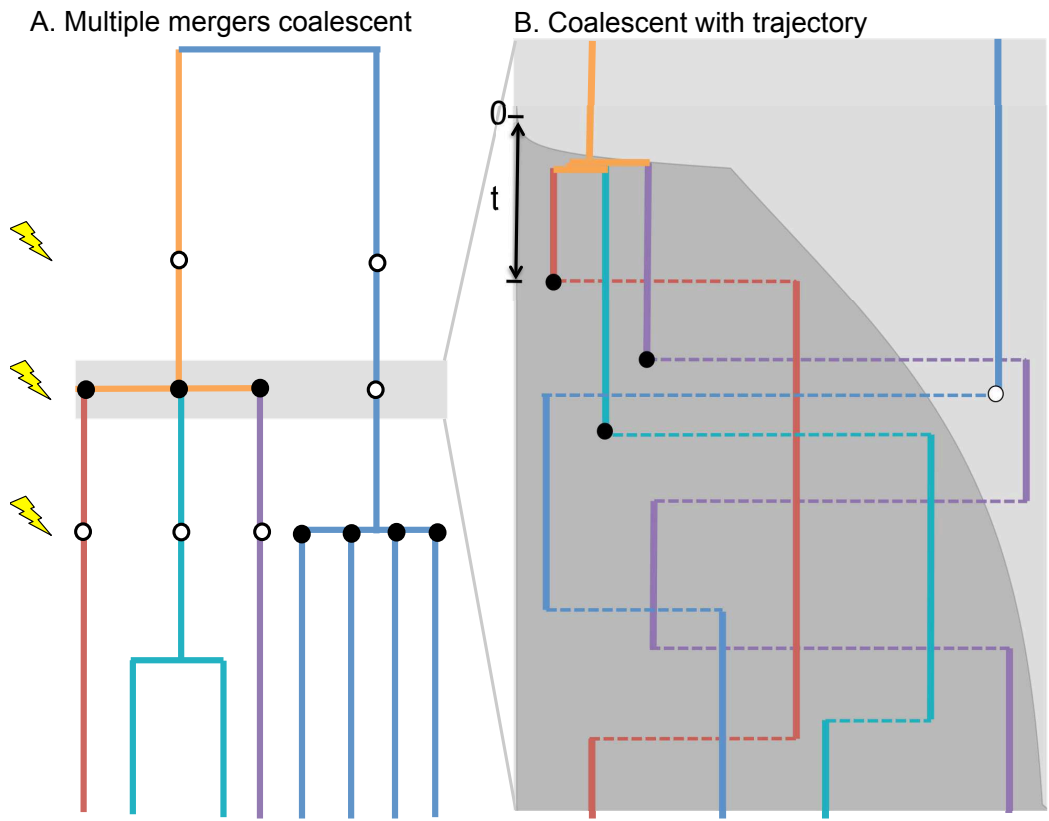}
\caption{ {\bf (A) An example of a multiple-merger coalescent genealogy.} Eight alleles have been sampled in the present day, and we trace their lineages backwards through time, up the page.  Lightning bolts indicate the times when a selected allele has swept into the population.  
At each sweep, each lineage is either descended from the original carrier of the derived allele at the selected site (lineages marked with a black dot) or from some other ancestor (lineages marked with a white dot). 
{\bf (B) Zooming in on one sweep.} The frequency of the derived allele, $D$, through time, $X(t)$, is shown in dark grey. The four surviving lineages are shown in different colors as in (A). Horizontal dotted lines depict recombination events in the history of a lineage. A dot indicates the oldest recombination event experienced by each of our lineages before the $D$ allele arose, and the color of the dot indicates where the allele recombined onto the $D$ background (black) or on to the $A$ background (white). As we approach the time the selected allele arose, the three lineages found on the $D$ background coalesce into a single lineage. 
}\label{cartoon_sweep}
\end{figure}

Figure \ref{cartoon_sweep}A shows an example of the relationships between different sampled individuals
at a neutral locus
in a finite population undergoing recurrent selective sweeps.
% coalescent process at a neutral locus under the coalescent process that approximates a model with recurrent selective sweeps and genetic drift. The coalescent event closest to the present day and the most ancient coalescent event are due to the finite population size, and so can only involve a pair of lineages. 
At the times indicated by the lightning bolts, selective alleles sweep into the population at some locus linked to our neutral site. 
All lineages descended from the original carrier of the derived allele coalesce,
nearly instantaneously on this time scale.
%while the second class corresponds to coalescences between lineages who are all descended from the original carrier of the derived allele.

Figure \ref{cartoon_sweep}B zooms in on one of these selective sweeps.  The derived allele at the selected locus ($D$) arose $\tau$ generations ago. The 
 five surviving ancestral lineages recombine on and off the $D$ background, whose frequency through time is shown by the dark grey shading. Just after time 0 those lineages on the $D$ background coalesce as $X$ goes to zero (their coalescent rate, which is proportional to $1/X$, goes to infinity). 
 We will show that the complexity of the process shown in Figure \ref{cartoon_sweep}B can be approximated by a much simpler multiple merger coalescent process suggested by Figure \ref{cartoon_sweep}A,
in which lineages coalesce ``neutrally'' at rate $1/(2N)$, 
and furthermore, each lineage flips a coin at each selective sweep to decide which type they are, 
and those that are of type $D$ merge simultaneously.

Suppose that a derived allele at the selected locus ($D$) arose $\tau$ generations ago, at time $0$.
The selected mutation may still segregate within the population in the present day, 
or may have gone to fixation or loss sometime before the present (in which case $X(\tau)=1$ or 0 respectively). 
First consider coalescences occurring very close to the origin of a selective mutation.
A lineage can be type $D$ at time $0$ for one of two reasons: 
either it was of type $D$ in the present day and not yet recombined off the $D$ background, 
or at the first recombination after the selected allele arose, the lineage chose to be of type $D$. 
The lineage of an individual drawn at random from the present-day population is therefore of type $D$ at time $0$ with probability
\begin{equation} 
  q = q(r,X) := X(\tau) e^{-r \tau } + r \int_0^{\tau} e^{-rt} X(t) dt.
\end{equation}
Here the integral is over $t$, the number of generations between the origin of $D$ and the first subsequent recombination on a lineage
($t$ is marked for the red lineage in Figure \ref{cartoon_sweep}B).
Note that although many recombination events may have occurred, 
since at each recombination event the lineage chooses a new type independently of its previous type,
we need only consider the first after the sweep.
If $\tau$ is much larger than $1/r$ the first term can be ignored, so we commonly assume that
\begin{equation} 
q(r,X) =  r \int_0^{\infty} e^{-rt} X(t) dt \label{general_q},
\end{equation}
as the allelic state of the sample has long been forgotten. Importantly, we can see that the dependence of $q$ on $X$ decays exponentially through time at rate $r$. 
Therefore, the fate of the selected allele more than a few multiples of $r$ after it arose, including its presence or absence in the present day, will have little effect on $q$.
Concretely, for two trajectories labeled 1 and 2, if $X_1(s) = X_2(s)$ for all $0 \le s \le T$, 
then regardless of subsequent differences in the trajectories, 
$|q_1-q_2| \le e^{-rT}$.

We can now approximate the rapid coalescence of lineages that are forced by the sweep by assuming that all lineages descended from the original carrier of the $D$ allele coalesce {\em simultaneously} when the selected allele appears (a ``multiple merger''). 
The lineages will actually coalesce at slightly different times, 
but the assumption the derived allele increases rapidly implies that this difference is small on the coalescent time scale $(i.e.\ o(2N))$. 
As each lineage takes part in this merger independently with probability $q$, 
the probability that $i$ out of $k$ surviving lineages coalesce at time $0$ is
\begin{equation}
  {k \choose i} q^{i}(1-q)^{k-i}, \quad \mbox{for} \quad 2 \leq i \leq k ,\label{q_binom}
\end{equation} 
reducing the number of lineages from $k$ to $k-i+1$.  

This approximation assumes that each lineage makes an independent choice of whether to recombine off the sweep, 
which is equivalent to assuming that the coalescences caused by the sweep form a `star'-like tree, with no internal edges of nonzero length.
Therefore, the approximation ignores dependencies between lineages induced by coalescent events earlier in the sweep, and so is a poorer approximation for large number of lineages. 
More sophisticated approximations have been developed to account for this dependency, which improve the properties for large samples \citep{Barton:98, Durrett:04, Etheridge:06,Pfaffelhuber:06}.
However, we believe this approximation captures many of the important features.
% To keep the results intuitive, we will use this simple approximation, but note that some features of these better approximations could be incorporated into this framework.

The other component of our approximation is that at all time, all pairs of lineages coalesce at rate $1/(2N)$ regardless of their allelic background. This approximation ignores the fact that lineages that are currently on different backgrounds cannot coalesce and that lineages on the same background coalesce at a higher rate (see equation \eqref{allelic_coal_rate}).  

We should also note that although large changes in the allele frequency over a small number of generations represent a large number of children
descended from a smaller number of ancestors, this will not cause rapid coalescence in a large population if the allele remains at intermediate frequencies. 
Concretely, consider a short time interval from generation $t_1$ to generation $t_2$, over which interval $X(t) \gg (t_2-t_1)/N$. The chance that any coalescence occurs during this time interval on the derived background is small ($O((t_2-t_1)/(X(t)N))$),  regardless of how the frequency $X$ changes. Therefore, large, sudden changes in allele frequencies will only force coalescence on the derived background if $X(t)$ is of order $1/N$ (and similarly for the ancestral background). For sites that are only partially linked to the selected locus, if recombination is moving the lineages across backgrounds at a sufficiently high rate compared to neutral coalescent rate ($Nr \gg 1$), then two lineages in this subdivided model coalesce at a rate close to $1/2N$ (see \citet{Hudson:88, Hey:91,Nordborg:97}, and \citet{Barton:04} for a detailed discussion). As such our approximation will therefore be worse close to the selected site, but is asymptotically correct for large $r$.

%Given all of the above considerations, the long term behavior of the selected allele does not matter to the coalescent process at sites that are only partially linked to the selected site. The coalescent process at such loci is almost entirely determined by the initial behavior of selected alleles after they enter the population,
%over a time scale given by $r$.

%%%%%%
\subsubsection{A simple trajectory} \label{ss:simple_trajectory}

To build intuition, we first consider a simple trajectory, making further approximations to keep the results accessible,
and compare the results to full coalescent simulations.
Assume that $D$ arises $\tau$ generations ago at a site at distance $r$ from the neutral site under consideration, 
rapidly sweeps to frequency $x$,
and remains close to this frequency for a time much greater than $1/r$. 
Under many models of directional selection, most of the time spent in reaching $x$ is spent at low frequency, 
so that any recombination that occurs during this time will likely move a lineage to the ancestral type,
and so only lineages that do not recombine during the initial sweep will coalesce.  
If we let $t_x$ be the time it takes for the selected allele to sweep to $x$
and assume $r\tau \gg 1$,
then a simple approximation to $q(r,X)$ is therefore 
(with the subscript emphasizing dependence on $x$)
\begin{equation} 
q(r,X) \approx q_x := x e^{-r t_x}. \label{eqn:qgoestox}
\end{equation}
If the initial increase of $D$ is driven by additive selection of strength $s$ with $Ns>1$, 
then the initial trajectory of $D$ will be logistic, and
it is reasonable to take $t_x= \log\big(\alpha  x/(1-x) \big )/s$, 
where $\alpha$ is $2N$ or $4Ns$ depending on whether $s$ is of order $1$ or $1/N$ 
the latter case corresponding to the case where  the selected allele has to rapidly achieve frequency $1/(Ns)$ to escape loss by drift).
Using $q_x$ to approximate the probability that a lineage is caught by the sweep, the expected pairwise coalescent time is smaller by a factor of
\begin{equation}
(1 - q_x^2 e^{-\tau / (2N)}) \label{mean_coal}
\end{equation} 
which can be found by considering whether a pair of lineages coalesce before, during, or after the sweep. 

If rather than remaining near $x$, the selected allele continues to sweep to fixation
-- perhaps it is still under selection with strength $s_2 \gg r$ --
then $q_x \approx e^{-r t_x}$ because the selected allele has gone quickly to fixation as in a full sweep, and the only time for recombination is in the early phase of the trajectory $t_x$. 
On the other hand, if the allele became strongly deleterious ($-s_2 \gg r$), then
$q \approx 0$, because there is little chance of it contributing genetic material to the population. 
However, if selection subsequently experienced by $D$ is weak ($|s_2| \ll r$), 
so that subsequent dynamics of the selected allele are sufficiently slow, 
then $q$ and therefore the coalescent process are independent of the eventual fate of the selected allele. 
In summary, for $q_x$ to be a good approximation to $q(r,X)$ and for the sweep to have an appreciable effect on the coalescent, 
we need $|s_2| \ll r < s$.

\subsubsection*{Comparison to simulation}
\begin{figure} 
\includegraphics[width = \textwidth]{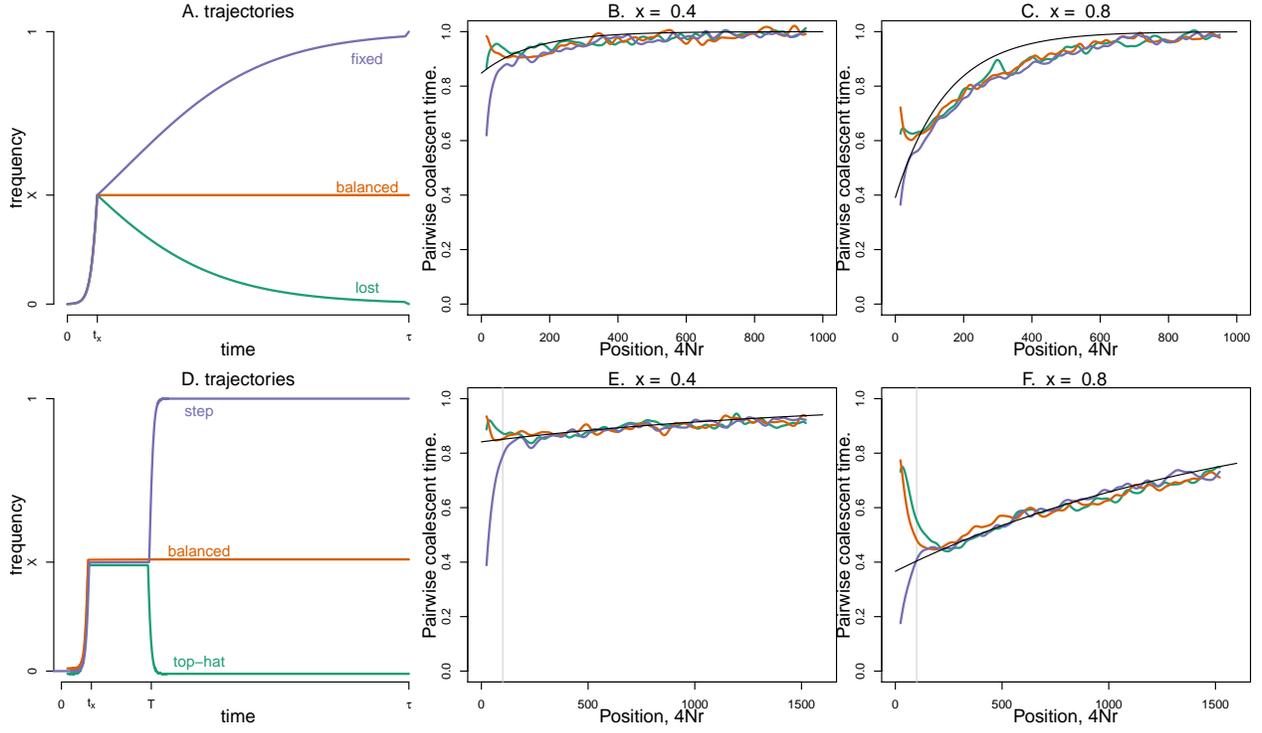}
\caption{{\bf The effect of a single partial sweep.} {\bf (A)} Three possible trajectories followed by the D allele after it arises $\tau$ generations ago, described in the text: blue is ``fixed'', green is ``lost'', and orange is ``balanced''.  {\bf (B)} and {\bf (C)} Mean pairwise coalescent time against recombination distance away from a selected site that has experienced one of the three types of sweeps shown in (A), with $x =0.4$ and $0.8$ respectively.  The other parameters were $t_x /2N = 6.6 \times 10^{-3}$ and $\tau /2N = 0.05$. {\bf (D)} Another 3 possible trajectories: green is ``top--hat'' and blue is ``step''. {\bf (E)} and {\bf (F)} Pairwise coalescent time as in (B) and (C), but using the trajectories shown in (D).  The other parameters were $t_x /2N =6.1 \times 10^{-4}$, $\tau/2N=0.1$ and $T/2N=0.02$.  The black line shows the approximation to the pairwise coalescent time of equation \eqref{mean_coal}. In E and F, the vertical line grey line marks $r=1/T$. %$4Ns=3000$.
\label{3_diff_sweeps}
}
\end{figure}

To demonstrate this,
we will apply the same approximation to situations with different long-term behaviors.
We consider five different possible trajectory types.
In all cases, the initial rise of $D$ was modeled as deterministic logistic growth begun at frequency $1/2N$ and adjusted to reach frequency $x$ after $t_x$ units of time.
In the first case (``balanced''), the allele remains thereafter at frequency $x$.
In the next two cases (Figures \ref{3_diff_sweeps}A--C), after time $t_x$, allele $D$ approaches either frequency 1 (``fixed'') or frequency 0 (``lost'') logistically,
reaching frequency $1-1/2N$ (or $1/2N$ respectively) after the next $\tau$ time units.
In the last two cases (Figures \ref{3_diff_sweeps}D--F), the allele $D$ 
remains at $x$ for $T$ generations, and then proceeds logistically, 
in time $t_x$, either to frequency $1-1/2N$ (``step'') or frequency $1/2N$ (``top-hat'').

In each case, 
we used {\tt mssel} (a modified version of ms \citep{Hudson:02} that allows an arbitrary trajectory, kindly supplied by Richard Hudson)
to simulate genealogies for a recombining sequence surrounding a selected locus at which a selected allele performs one of the trajectories shown in Figure \ref{3_diff_sweeps} . 
The average pairwise coalescence time from these simulations was calculated by dividing the pairwise genetic diversity by the mutation rate, and is shown in Figure \ref{3_diff_sweeps} at different distances from the selected locus,
compared to the quantity predicted by equation \eqref{mean_coal}.
Close to the selected site (e.g.\  for $r< 1/T$ in Figure \ref{3_diff_sweeps}E and F) the curves diverge, 
since the sites represented by the blue curves see a full sweep, reducing diversity close to the selected site, 
while those in the orange curves see a short-term balanced polymorphism, and hence show a peak in polymorphism near the selected site). 
As we increase recombination distance away from the selected site, the three curves are in good agreement with the black line (equation \eqref{mean_coal}), 
indicating that our partial sweep model captures the main effect on diversity. 

Our simple approximation describes diversity levels well at partially linked sites over a range of different scenarios, and works well for a wider range of parameters (results not shown). We furthermore used equation \eqref{q_binom} to predict the effect of this simple partial sweep on the coalescent process of more than two lineages, and found close agreement with further {\tt mssel} simulations for various summaries of diversity such as the expected number of segregating sites (results not shown).  Overall, these results confirm that for partially linked sites, the coalescent process is mostly determined by the initial rapid behavior of the selected allele.

\subsection{A recurrent sweep coalescent model}
We now consider patterns of diversity at a neutral locus 
affected by many different selected alleles that sweep into the population 
at the times of a homogeneous Poisson process with rate $\nu$.
We assume that the sweep rate is low enough that sweeps do not interfere with each other, 
and return to discuss this assumption later.
Each sweep occurs at some distance $r$ from the neutral locus, 
and as it sweeps its frequency follows some particular trajectory $X(t)$,
which together in equation \eqref{general_q} determine $q$, the probability that a lineage at the neutral site is caught by the sweep. 
Rather than try to explicitly model randomness in these two components,
at first we will assume that each sweep independently chooses its value of $q$ from a probability distribution with density $f(q)$. 
This model is exactly a Lambda coalescent, with $\Lambda(dq) = q^2 \nu f(q) dq + \delta_0(dq)/2N$ \citep[see][for a recent review]{Berestycki2009},
but we leave our discussion in terms of $f$ to make the results more intuitive.

Following from our assumption that each lineage is affected by a given sweep independently with probability $q$,
when there are $k$ surviving lineages, the rate at which they coalesce to $k-i+1$ lineages due to sweeps is 
\begin{equation}
 \nu {k \choose i} \int_0^1 q^i(1-q)^{k-i} f(q) dq \label{k_to_i} .
\end{equation}
This follows from our assumption that sweeps occur homogeneously through time and do not interfere with each other,
and properties of marked Poisson processes.
For ease of presentation we denote
\begin{equation}
  I_{k,i} =  {k \choose i} \int_0^1 q^i(1-q)^{k-i} f(q) dq  . \label{I_defn}
\end{equation}
Recall that under our model, the rate of coalescence of pairs of lineages due to genetic drift is $1/(2N)$,
so that the rate at which the coalescent process with $k$ lineages coalesces to $k-i+1$ lineages is 
\begin{equation}
  \lambda_{k,i} = {k \choose 2} \frac{1}{2N} \delta_{i,2} +  \nu I_{k,i} \quad \mbox{for} \; 2 \leq i \leq k \label{k_to_i_w_drift} ,
\end{equation} 
where $\delta_{i,2} =1$ if $i=2$ and $0$ otherwise.
The total rate of coalescent events when there are $k$ lineages is therefore
\begin{equation}
  \lambda_{k} = \frac{1}{2N} {k \choose 2} +  \nu \sum_{i=2}^k I_{k,i}  \quad \mbox{for} \; k \ge 2, \label{tot_lambda}
\end{equation}
and conditional on a coalescent event the probability that $i$ lineages out of $k$ coalesce, 
reducing from $k$ to $ k-i+1$ lineages, 
is 
\begin{equation} \label{choose_event}
  p_{k,k-i+1} =\frac{\lambda_{k,i}}{\lambda_{k}} = \frac{ \frac{1}{2N} {k \choose 2} \delta_{i,2} + \nu I_{k,i} }{ \frac{1}{2N} {k \choose 2} + \nu \sum_{i=2}^k I_{k,i} }, \quad \mbox{for} \; 2 \leq i \leq k . 
\end{equation}
To simulate from this coalescent process we can simulate an exponential waiting time with rate $\lambda_{k}$, 
pick a number of lineages to coalesce using probabilities $p_{k,k-i+1}$, and run this process until we have a single lineage remaining.

Note that in deriving this process we have assumed that at all times, lineages also coalesce at a neutral rate $1/2N$. 
This can be justified by assuming that recombination moves lineages between backgrounds at a high enough rate 
to allow the effects of the partitioning of the population by segregating alleles to be ignored. 
Therefore, the approximation will break down if a typical neutral site, at any given time, is close enough  (e.g.\ within an $r$ of order $1/N$) to an allele 
maintained at intermediate frequency by long-term balancing selection (e.g.\  alleles maintained for time scales of order $N$). 
Further work is needed to refine the coalescent under those conditions, but our approximations should be suitable for a broad range of scenarios and genomic regions.

\subsection{The coalescent process with homogeneous sweeps} 
It is natural to examine the case in which selective sweeps occur at uniform rate along a sequence of total length $L$. 
We assume that this sequence recombines at rate $r_{BP}$ per base each generation, 
and that sweeps enter the population at a rate $\nu_{BP}$ per base each generation, 
so that the total rate of sweeps is $\nu = \nu_{BP} L$.  
We also assume that the sweeps are homogeneous, 
i.e.\ the trajectory followed by the frequency of the derived allele, $X$,  
is independent of the distance between our neutral site and the site at which a sweep occurs.

We will consider sweeps occurring along a very long chromosome and so will take $L \to \infty$,
but then the total rate of sweeps, $\nu = \nu_{BP} L$, also goes to infinity.
To obtain a meaningful limit, we need that as $L \to \infty$
the rate of sweeps corresponding to each nonzero value of $q$ converges to a finite value. 
Recall from \eqref{general_q} that the probability a lineage is caught up in a given sweep 
depends on the distance to the sweep (which is $r_{BP}\ell$ for a site $\ell$ bases away) 
and the trajectory $X$ taken by the sweep, and is given by $q(r_{BP}\ell,X) = r_{BP}\ell \int_0^\tau \exp(-r_{BP}\ell t) X(t) dt$.
In a finite genome of length $L$, the probability distribution on values of $q$ has density
$f(q) = h_L(q)/L$, where $h_L(q) = \int_0^L \P_X\{q(r_{BP}\ell,X) \in dq\} d\ell$. 
Here $h_L(q)$ is the rate at which selective sweeps appear at location $r_{BP}\ell$ 
and whose trajectory $X$ gives $q(r_{BP}\ell,X)=q$, integrated across the genome;
and $f(q)$ is $h_L(q)$ normalized to integrate to 1, since $\int_0^1 h_L(q) dq = L$.
The functions $h_L$ converge for $q>0$ as $L$ becomes large 
as long as the probability that distant sweeps affect the focal site decays quickly enough.
We therefore assume that $h_L(q)$ converges to a finite limit $h(q)$, i.e.
that the following exists:
\begin{equation}
  h(q) = \lim_{L \to \infty}  L f(q) ~~\quad \mbox{for} \; 0 < q \le 1 . \label{homogen_limit}
\end{equation}
This means that although the total rate of sweeps per generation is infinite, 
only a finite number happen close enough to our neutral site to potentially affect our coalescent process. 
Therefore, the rate at which $k$ lineages coalesce down to $k-i+1$ due to sweeps converges:
\begin{equation}
\nu_{BP} \, L \,  I_{k,i} \rightarrow \nu_{BP} {k \choose i} \int_0^1 q^i(1-q)^{k-i} h(q) \, dq \quad ~~\mbox{as} \;~~ L \to \infty . \label{homo_I}
\end{equation}

If we take the trajectory $X$ to be fixed, we can rewrite equation \eqref{homo_I} as
\begin{align}
 \nu_{BP}  {k \choose i}\int_0^1 q^i(1-q)^{k-i} h(q) dq &= \nu_{BP}  {k \choose i}\int_0^{\infty} q( r_{BP} \ell,X)^i(1- q( r_{BP} \ell,X))^{k-i} d\ell \nonumber \\
&= \frac{\nu_{BP}}{r_{BP}}  {k \choose i} \int_0^{\infty} q(r,X)^i(1-q(r,X))^{k-i} dr , \label{rate_2}
\end{align}
which decouples the dependency of the rate of sweeps on the recombination rate $r_{BP}$ from the trajectory $X$.
If $X$ is random, then we need to average over possible trajectories, and so we define
\begin{equation}
  J_{k,i} = {k \choose i} \E_X \left[ \int_0^\infty q(r,X)^i(1-q(r,X))^{k-i} dr \right] ,
\end{equation}
where $\E_X[\cdot]$ denotes the average over possible trajectories. 
We will assume that this integral is finite for $2 \le i \le k$; for further discussion of these points see Appendix \ref{I_generalized}. 
Importantly, under our assumption that sweeps do not interfere with each other,
$J_{k,i}$ does not depend on the recombination rate $r_{BP}$ or the rate of sweeps $\nu_{BP}$, 
but only on the dynamics of the selective sweeps $X$. 

Allowing coalescent events due to drift, $k$ lineages coalesce down to $k-i+1$ at rate 
\begin{equation} \label{eqn:J_rates} 
  \lambda_{k,i} = \frac{1}{2N} {k \choose 2} \delta_{i,2} + \frac{\nu_{BP}}{r_{BP}} J_{k,i}   \quad \mbox{for} \; 2 \leq i \leq k ,
\end{equation} 
where $\delta_{i,2}=1$ if $i=2$ and is 0 otherwise.
As equation \eqref{eqn:J_rates} follows from the simple change of variable in equation \eqref{rate_2} it will hold under any homogeneous sweep model where sweeps instantaneously (relative to a time scale of $2N$) force lineages to coalescence, 
with $J_{k,i}$ replaced by some constant that does not depend on $r_{BP}$ or $\nu_{BP}$. 
This result greatly generalizes that of \citet{Kaplan:89} who described a similar coalescent process for a full sweep model. 

We can see from equation \eqref{eqn:J_rates} that $2N \nu_{BP} / r_{BP}$ is the relevant compound parameter that in a general sweep model determines the rate of sweeps relative to neutral coalescent events. In small samples, sweep-induced coalescent events will dominate those due to drift if the population-scaled rate of sweeps per unit of the genetic map is much greater than one, provided that not all the $J_{k,i}$ are too small. We revisit this strong sweep limit in Section \ref{strong_limit}.

\subsubsection*{The coalescent process with homogeneous partial sweeps.}
We now return to the setting of section \ref{ss:simple_trajectory}, in which a simple trajectory rises quickly to frequency $x$,
under which assumptions $q(r,X)\approx q_x$ (equation \eqref{eqn:qgoestox}).
We suppose that the frequency $x$ at which each sweep slows is chosen independently
with probability density $g(x)$. It also seems reasonable to assume furthermore that $t_x$, the time it takes to reach frequency $x$, does not depend on $x$; we will denote this time by $t$. This is approximately true for many models of directional selection, since selected alleles move quickly through intermediate frequencies. In this case, the rate at which $k$ lineages coalesce to $k-i+1$ is
\begin{equation}
\lambda_{k,i} \frac{1}{2N} {k \choose 2}\delta_{i,2} + {k \choose i} \frac{\nu_{BP}}{t\, r_{BP}} \int_0^{\infty} \left( \int_0^1 \left(xe^{-r} \right)^{i}\left(1-xe^{-r}\right)^{k-i} g(x) dx \right) dr, \label{lambda_partial_homogen} 
\end{equation} 
suggesting that the important quantity, which acts as a coalescent time scaling, is $2N\nu_{BP}/(t\, r_{BP})$, with the distribution on $x$ acting to control how many lineages are forced to coalesce with each sweep. 
If we determine $t$ by a simple model of additive selection with selection coefficient $s$, the key parameter becomes $2N\nu_{BP} s/(\log(Ns)\, r_{BP})$.

This compound parameter,  $2N\nu_{BP} s/(\log(Ns)\, r_{BP})$, is also the key parameter in full sweep models \citep{Kaplan:89, Stephan:92}. 
However, since full sweeps require $x=1$,
if diversity is strongly reduced then numerous lineages must merge at each sweep,
which in turn leads to a strong skew towards rare alleles in the frequency spectrum. 
We will see that this relationship between the reduction in diversity 
and the skew in the frequency spectrum is substantially weakened under a partial sweep model when we allow $x \ll 1$.

\subsection{Summaries of neutral genetic diversity.}

\subsubsection{Level of neutral diversity.}
A key quantity of interest is the level of neutral nucleotide diversity, $\pi$, the number of differences between randomly sampled alleles at a neutral locus. 
Under an infinite sites model of mutation, which we will use here, 
the expectation of $\pi$, averaging across sites,
is equal to the expected coalescent time of a pair of lineages multiplied by twice the mutation rate.  
If the mutation rate per generation at our neutral locus is $\mu$, in the absence of sweeps, the level of diversity is $\E[ \pi] = \theta $,
where $\theta=4N \mu$ is the population-size scaled mutation rate, and the expectation is the average across sites.
Note that $\theta$ is the level of diversity under the usual neutral model.

Under our model featuring both sweeps and drift,
\begin{equation}
\E[\pi] = \frac{\theta}{1 + 2N I_{2,2} \nu} .
\end{equation}
so a key parameter is the population--scaled rate of sweeps $2N\nu$.

\begin{figure}
\includegraphics[width =\textwidth]{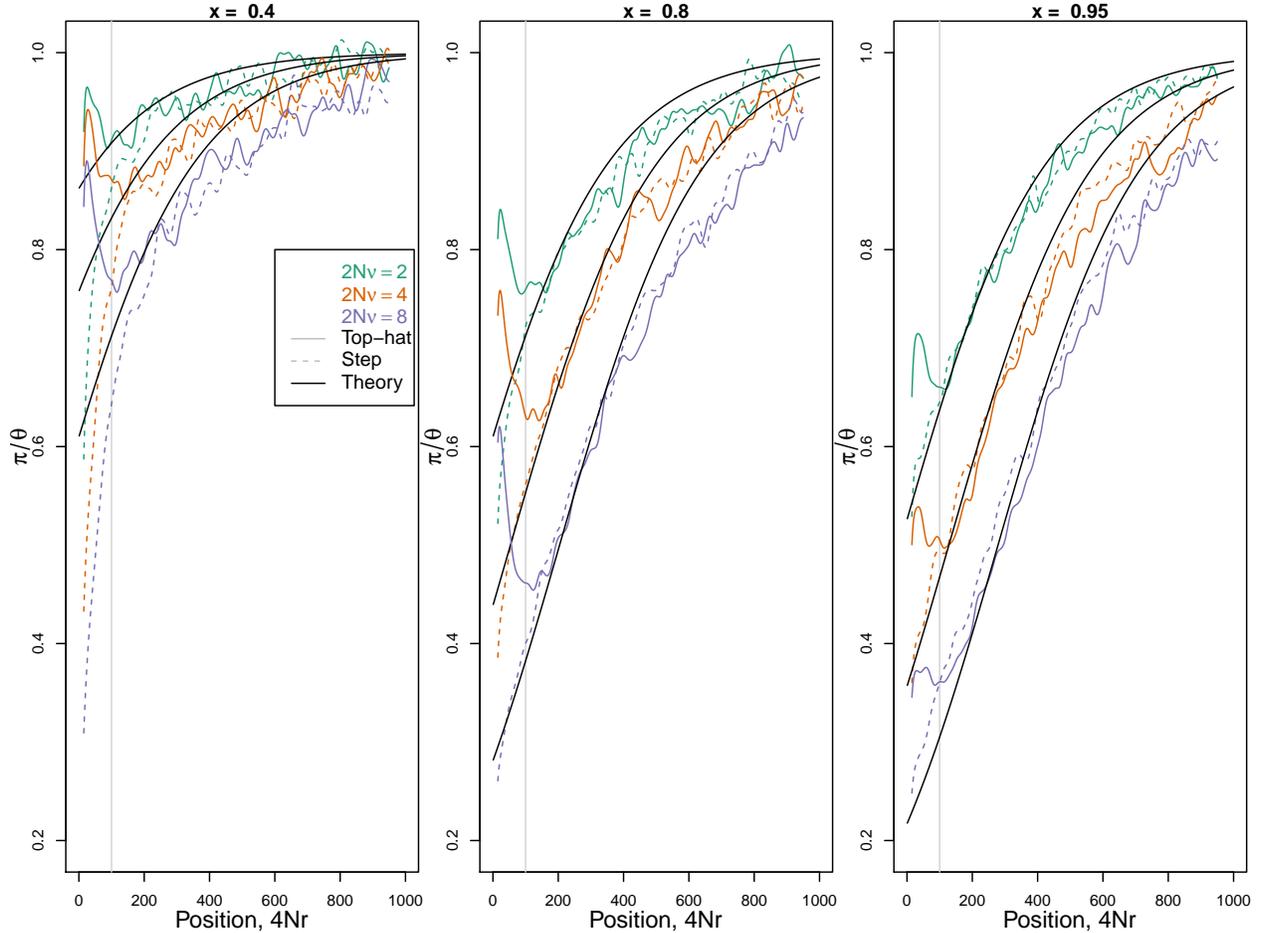}
\caption{ {\bf Reduction in diversity ($\pi/\theta$) as a function of recombination distance} from a site experiencing recurrent sweeps. The three panels are for different values of the frequency $x$ that each sweep reached rapidly.  
The solid line is for recurrent top-hat trajectories
and the broken line for recurrent step trajectories 
The time that the trajectory pauses is $T/2N=0.01$ and $t_x/2N=0.003$ in both cases. 
The three colors correspond to three different population-scaled rates of sweeps: $2N\nu=$ $2$, $4$ and $8$.
The vertical grey line marks recombination distance $r>1/T$ from the selected locus, above which the dynamics subsequent to reach $x$ should make little difference. The solid black lines give the prediction of \eqref{recurr_pi_fixed_site}. }
\label{recurr_pi}
\end{figure}

To examine the applicability of our approximations we again performed coalescent simulations with {\tt mssel} 
for a selected locus at a fixed location experiencing recurrent sweeps. 
In this case, where selected alleles recurrently sweep into the population at a \emph{fixed} genetic distance $r$, 
following our simple partial sweep trajectory again as characterized by $q_x$ and $2N$, 
the nucleotide diversity is given by
\begin{equation}
\E[\pi] = \frac{ \theta }{1 + 2N\nu x^2  \exp\left(-2 r t_x \right) }. \label{recurr_pi_fixed_site}
\end{equation}
We used two types of recurrent trajectory --
the recurrent `step' and the recurrent `top-hat', as described earlier.
For the recurrent top-hat trajectory, we simulated an exponential waiting time with mean $\nu$ between the end of one `top-hat' and the start of the next 
(and similarly for the `step' case). 
In Figure \ref{recurr_pi} we show diversity levels moving away from the locus undergoing these two types of recurrent sweeps, as well as the analytical approximation given by equation \eqref{recurr_pi_fixed_site}. Recall that in both types of trajectories the derived allele pauses at frequency $x$ for time $T$, 
and therefore we expect that the fate of the allele will affect diversity at recombination distances smaller than $1/T$. 
For distances larger than $1/T$, equation \eqref{recurr_pi_fixed_site} shows good agreement with our simulations, regardless of whether the recurrent sweeps go to loss or fixation. 
The approximation does not perfectly match our simulations, presumably because $e^{-r2t_x}$ is an imperfect approximation to the probability of recombination during the sweep. Nevertheless, diversity levels generated by the two types of recurrent trajectory agree away from the selected site, 
which importantly confirms that only the initial rapid stage of the trajectory affects the coalescent process at partially linked sites. 

%The key quantity here is $2N \nu_{BP} /r_{BP}$, the population scaled rate of sweeps per genetic map unit,
%and linked selection will cause much more coalescence than will drift
%when this quantity is much larger than one.

\paragraph{The level of diversity under homogeneous sweeps.} Under the model in which sweeps occur homogeneously along an infinite sequence, 
with coalescent rates given by equation \eqref{eqn:J_rates}, the level of nucleotide diversity is given by
\begin{equation}
  \E[\pi] = \frac{ \theta }{ 2N\nu_{BP} J_{2,2}/r_{BP} +  1 } . \label{pi_I}
\end{equation}
These results generalize previous results by \citet{Kaplan:89} and \citet{Stephan:92}, who found a relationship of the form \eqref{pi_I} for a model of homogeneous recurrent full sweeps. In fact, since equation \eqref{pi_I} follows only from 
the assumption that the rate and characteristics of sweeps are independent of their location along the genome (see equation \eqref{rate_2}),
this relationship between diversity, the density of selective targets, and recombination rate
will hold for a wide variety of homogeneous recurrent sweep models including the homogeneous full sweep model.

\subsubsection{Frequency Spectrum.}

We now study the effects of recurrent partial sweeps on other properties of neutral diversity at a locus 
besides pairwise nucleotide diversity, and compare our calculations to simulation.

Two commonly studied properties of a sample of neutral diversity at a locus are the expected number of segregating sites in a sample of size $n$,
and the expected number of singletons in a sample of size $n$. Under the infinite-sites assumption, these are respectively equal to the mutation rate multiplied by the expected total length of the genealogical tree of the sample (which we denote $T_{tot}$)  and by the mutation rate multiplied by the expected total length of the terminal branches ($T_1$). 
We provide recursions that allow easy calculation of both $\E[T_{tot}]$ and $\E[T_{1}]$ in Appendix \ref{expected_times_append}.

We also look more generally at the frequency spectrum of segregating alleles,
which is, in a sample of $n$ individuals, the proportion of segregating sites at which $k$ derived alleles are found,
for each $1\le k \le n$.
Let $F_{n,k}$ denote the expected proportion of segregating sites in a sample of size $n$ at which exactly $k$ samples carry the derived allele
under an infinite sites model of mutation.
$F_{n,k}$ is equal to the expected time in the coalescent tree spent on branches that subtend exactly $k$ tips
(those on which mutation would lead to a site segregating at $k$ out of the $n$ samples),
divided by $\E[T_{tot}]$.
Under neutrality (Kingman's coalescent), this quantity is $F^N_{n,k} = (1/k)/\sum_{j=1}^{n-1} (1/j)$.
It is not so easy to find an explicit general expression under the coalescent model with sweeps that we study,
but for the case $k=1$ we have described in Appendix \ref{expected_times_append} how to compute $\E[T_1]/\E[T_{tot}]$,
and the general case can be found from simulation of the coalescent process.

\begin{figure}  
  \includegraphics[width =\textwidth]{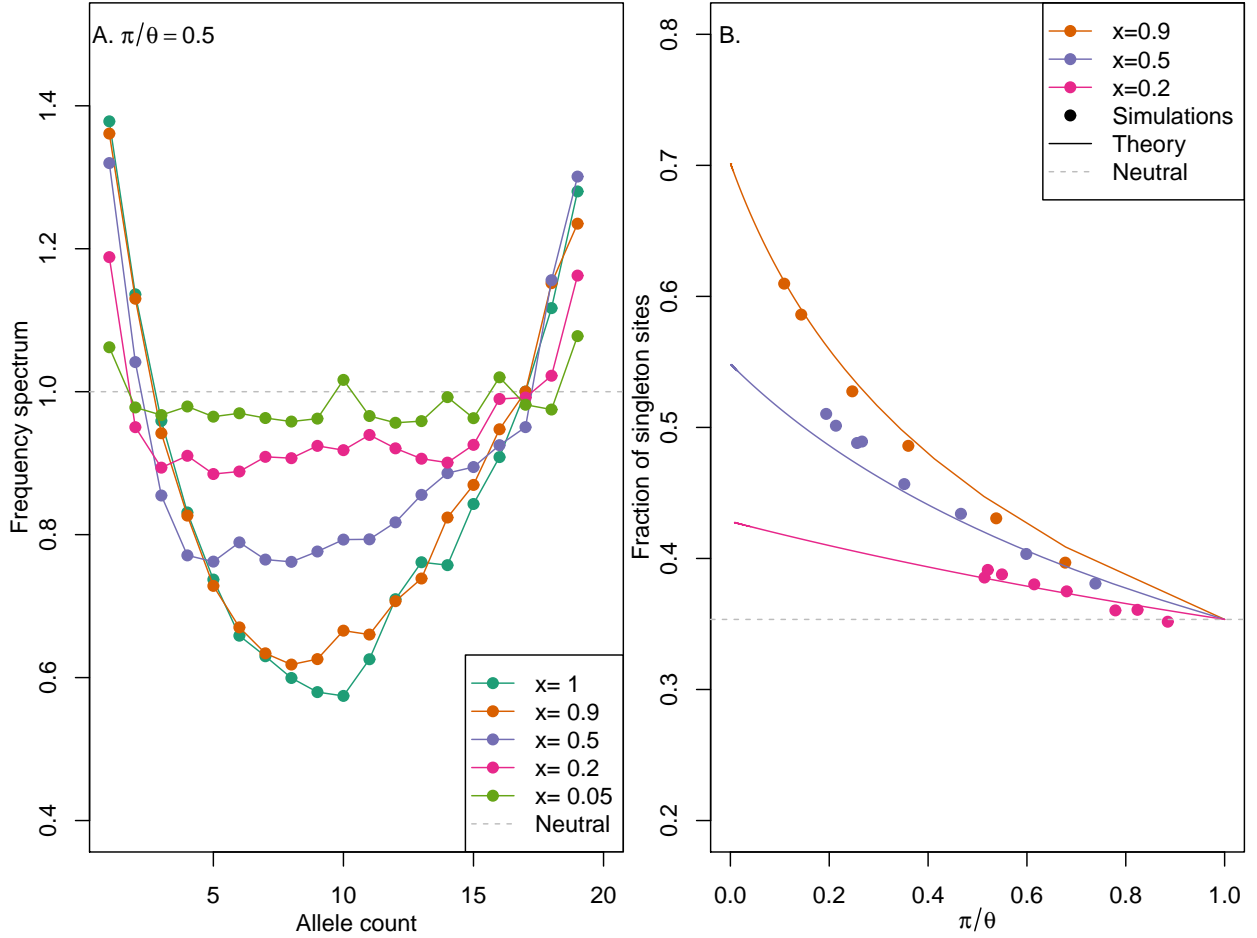}
  \caption{{\bf Properties of the frequency spectrum with sweeps occurring at a fixed genetic distance} 
  Coalescent rates are given by equation \eqref{k_to_i_w_drift}, with $q$ fixed to $q_x=x e^{-t_x r}$ and $t_x r =0.6$, across a range of $x$. 
  {\bf (A)} The percentage of segregating sites found at frequency $1\le k \le 20$, relative to the neutral expectation (i.e.\ $F_{20,k}/F^N_{20,k}$). 
  In these simulations the rate of sweeps $N\nu$ has been fixed to result in a $50\%$ reduction in diversity. 
  The dotted grey line gives the neutral expectation.  
  {\bf (B)} The mean number of singletons divided by mean number of segregating sites, from {\tt mssel} simulations with a sample size of $10$ at a neutral site a distance $2Nr=200$ from a selected site.
  The selected allele performs a recurrent top-hat trajectory (with $N=10,000$ and $t_x/2N=.003$, giving $rt_x= 0.6$, and pausing $T/2N=0.01$) to frequency $x=0.2$, $x=0.5$, or $x=0.9$ across a range of $2N\nu$. 
  Note the span of $\pi/\theta$ is smaller in the low $x$ simulations as the effect on diversity of a given $2N\nu$ is smaller. 
  Solid lines show the analytical approximation for $\E[T_{1}] / \E[T_{tot}]$ of Appendix \ref{expected_times_append}.
  The dotted grey line gives the neutral value of the expected proportion of singletons $1/\sum_{j=1}^{n-1} 1/j$. 
} \label{Freq_spec_fixed_q}
\end{figure}

Figure \ref{Freq_spec_fixed_q}A shows the ratio of $F_{n,k}/F^N_{n,k}$, estimated by direct simulation of our coalescent process. 
The rates are given by equation \eqref{k_to_i_w_drift}, with $q$ fixed to $q_x=x e^{-t_x r}$, 
% representing recurrent partial sweeps to $x$ at a fixed distance 
and $t_x r =0.6$ (and various $x$). 
To make the simulations comparable, the population scaled rate of sweeps $2N\nu$ was adjusted such that $\pi/ \theta=1/2$ in each, 
i.e.\ to obtain a $50\%$ reduction in diversity due to sweeps. 
We see that for partial sweeps at a fixed site, across a range of $x$, 
the frequency spectrum is skewed towards rare alleles and away from intermediate frequency alleles. 

To test the degree to which our coalescent matches the full model,
in Figure \ref{Freq_spec_fixed_q}B we compare the mean proportion of singleton sites under our coalescent model
to that found from simulation with {\tt mssel}.
We simulated a recurrent top-hat trajectory of the frequency at a selected locus as before, 
and used this trajectory with {\tt mssel} to simulate the neutral coalescent at a non-recombining locus a distance $r$ away from this selected locus.
We used the three values $x =0.9$, $0.5$, and $0.2$ for the intermediate frequency the allele reached,
and in each case varied the rate of sweeps, $\nu$
Each combination of $\nu$ and $x$ gives a point in Figure \ref{Freq_spec_fixed_q}B,
plotted at its mean reduction in diversity ($\pi/\theta$)
and the mean number of singletons divided by the mean number of segregating sites.
These are compared to the analytical values of $\E[T_1]/\E[T_{tot}]$ computed using equations \eqref{num_tot} and \eqref{num_single}, with coalescent rates given by equation \eqref{k_to_i_w_drift}, using a constant $q = x e^{-r t_x}$ and \eqref{pi_I} to find the reduction $\pi/\theta$.
There is good agreement between the simulations and the analytical results, 
showing that our simplified process approximates the properties of the full coalescent process at a single site reasonably well. 

\begin{figure}  
\includegraphics[width =\textwidth]{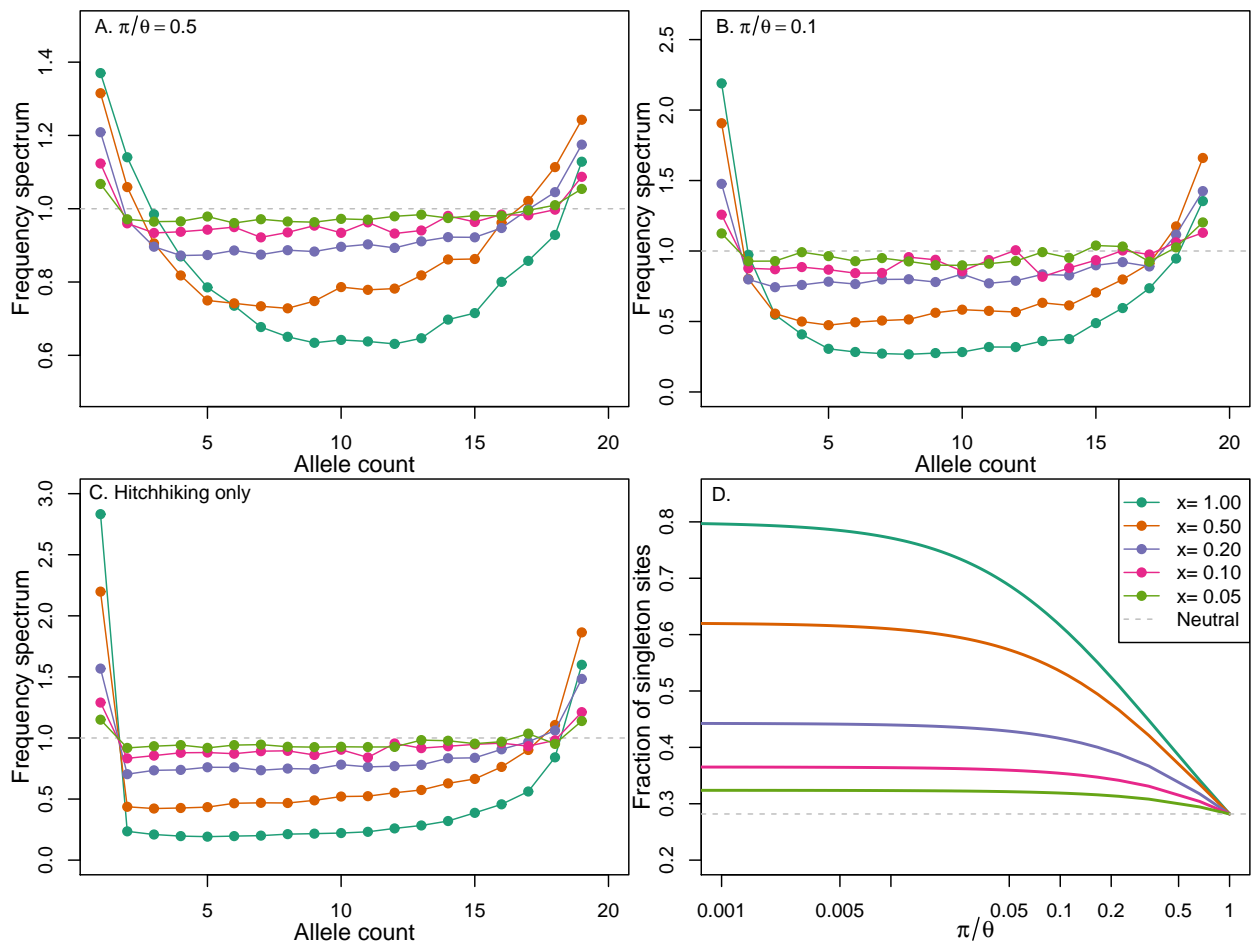}
\caption{{\bf Properties of the frequency spectrum under a spatially homogeneous model of sweeps} using the coalescent process with rates given by equation \eqref{lambda_partial_homogen}. Simulations were performed for a sample size of $20$. For a particular $x$ we adjusted the value of $N \nu_{BP} / (t r_{BP})$ to achieve the specified reduction in $\pi$. {\bf (A)} and {\bf (B)} The percentage of segregating sites found at frequency $1\le k \le 20$, relative to the neutral expectation for sweeps. In each panel the reduction in diversity, $\pi/\theta$ is fixed. {\bf (C)} The same quantities as in A and B, but for the case where there is no genetic drift, and sweeps are the only stochastic force affecting allele frequencies. {\bf (D)} The fraction of segregating sites that are singletons, for different $x$, as a function of $\pi/\theta$, calculated using recursions for $\E[T_{1}]/\E[T_{tot}]$ (Appendix \ref{expected_times_append}).  
\label{Freq_spec_const_x_homogen}
}
\end{figure}

Figure \ref{Freq_spec_fixed_q} studied the effect on the frequency spectrum of recurrent sweeps at a fixed distance from a neutral site;
in Figure \ref{Freq_spec_const_x_homogen} we study the frequency spectrum under the coalescent process with sweeps occurring homogeneously along the genome.
Figures \ref{Freq_spec_const_x_homogen}A and B show the same quantities as Figure \ref{Freq_spec_fixed_q}A, 
for simulations of the homogeneous partial sweep coalescent process with a fixed value of $x$, using rates given by equation \eqref{lambda_partial_homogen}, and $2N\nu_{BP}/(t r_{BP})$ chosen so that $\pi$ is 50\% and 10\% of its value under neutrality respectively. 
In Figure \ref{Freq_spec_const_x_homogen}C, 
there is no genetic drift and only sweeps force coalescence, i.e.\ $N=\infty$ and so we do not need to specify $2N\nu_{BP}/(t r_{BP})$ as it acts only as a time scaling. 
In \ref{Freq_spec_const_x_homogen}D we show our analytic calculation of $\E[T_{1}]/\E[T_{tot}]$ as a function of the reduction in $\pi$ caused by selective sweeps.

The skew in the frequency spectrum depends strongly on the frequency $x$ reached by the selected allele. Sweeps to low frequencies lead to a much smaller distortion for the same reduction in $\pi$. 
Therefore, the strong relationship between the reduction in $\pi$ and the skew in the frequency spectrum under a model of full sweeps
is much weaker if the sweeps do not go to fixation.

Intriguingly, sweeps that go to intermediate frequency can lead to a greater proportion of high frequency derived alleles than under a full sweep model. While a single, recent full sweep leads to high frequency derived alleles through hitchhiking \citep{Fay:00}, under a recurrent full sweep model these alleles are then fixed in the population by subsequent sweeps and drift \citep{Kim:06}, and therefore removed from the frequency spectrum. Further work would be needed to understand the intuition for the excess of high frequency derived alleles under a recurrent partial sweep model.

\paragraph{Summaries of the frequency spectrum}
In Figures \ref{Freq_spec_fixed_q} and \ref{Freq_spec_const_x_homogen}, we saw that
regardless of whether sweeps occur at a fixed distance from our neutral site or homogeneously along the sequence, as we increase the rate of sweeps the frequency spectrum becomes further skewed towards rare derived alleles at the expense of intermediate frequency alleles.
Here we provide evidence that this will hold for any set of parameter values.
Tajima's $D$ and Fu and Li's $D$ \citep{Tajima:89, FuLi:93} are two common ways of detecting deviations away from the frequency spectrum expected under a neutral model with a constant population size. 
Negative values of Tajima's $D$ can be thought of as indicating a deficit of intermediate frequency alleles,
and Fu and Li's $D$ indicates an excess of singleton alleles. 
\citet{Durrett:05} proved that in large samples, both of these summary statistics are negative under a multiple mergers coalescent model of full sweeps, 
as long as $\lambda_k$, the total coalescent rate when there are $k$ lineages, satisfies
% $\lambda_k$ given by equation \eqref{tot_lambda} satisfies  
\begin{equation}
\sum_{k=2}^{\infty} \left(\lambda_k - {k \choose 2} \right) \frac{\log(k)}{k^2} < \infty \label{durrett_condition} .
\end{equation}
See equation (4.5) in \citet{Durrett:05}. Informally, this condition requires that the total coalescent rate is not too much higher than the neutral coalescent rate when there are a large number of lineages. Their methods were not specific to their situation but hold for all multiple merger coalescent models satisfying equation \eqref{durrett_condition}. As above, we argued that a generalized sweep model can be approximated by a multiple merger coalescent, and therefore, it seems that reasonable generalized sweep models will, at least for large samples, have a frequency spectrum that is skewed towards singletons at the expense of intermediate frequency alleles (a notable exception is the `low frequency' limit we discuss below).

\subsection{Limiting processes}
Before we move to discuss the implications of these results for data analysis there are two limiting processes that merit our attention. The first is when the rate of sweeps is sufficiently high to dominate genetic drift as a source of stochasticity.  The second limit results when sweeps very rarely achieve high frequency in the population, in which case the resulting coalescent model is identical to the standard ``neutral'' coalescent, despite that fact that much of the stochasticity may be driven by sweeps.

\subsubsection*{The rapid sweep limit} 
\label{strong_limit}

A surprising conclusion from the homogeneous model and equation \eqref{eqn:J_rates}
is that if all coalescences come from ``selective'' events, 
then the frequency spectrum does not depend on the density of selective targets or on the recombination rate
(although the number of segregating sites certainly does). This effect can be seen in Figure \ref{Freq_spec_const_x_homogen}D as the fraction of singleton sites plateaus when the reduction in $\pi$ is large, i.e.\ when the population scaled rate of sweeps per unit of recombination is high, $\nu_{BP}/r_{BP} \gg 1/2N$. 
The easiest way to see this is to take $N \to \infty$ while keeping the rate of sweeps and their trajectory dynamics fixed,
so that in a sample of fixed size the coalescence rate from equation \eqref{eqn:J_rates} converges to $\lambda_{k,i} \to \nu_{BP}/r_{BP} J_{k,i}$,
where $J_{k,i}$ does not depend on $\nu_{BP}$, $r_{BP}$, or $N$. In this limit, $\nu_{BP}$ and $r_{BP}$ only affect the process by a time scaling, 
do not affect the transition probabilities of equation \eqref{choose_event}, and so do not affect the frequency spectrum. 
Diversity in this limit behaves as
\begin{equation}
\E[\pi]= \frac{2 \mu r_{BP}}{\nu_{BP} J_{2,2}} .  \label{inf_hetero} 
\end{equation}
(assuming, as usual, that $\mu$ is sufficiently small)
i.e.\  nucleotide diversity increases linearly with the recombination rate, if neither $\nu_{BP}$ or $J_{2,2}$ varies across recombination environments. Similar limits can also be derived by letting $N \to \infty$ under the more general coalescent process with rates given by equation \eqref{k_to_i}.

For this limit to be a reasonable approximation for a sample of size $k$ in a population of size $N$,
we need the rate of neutral coalescences to be much smaller than the rate of selective coalescences,
i.e.\ ${k \choose 2} \ll  N \nu_{BP}/r_{BP}  \sum_{i=2}^k J_{k,i}$.  
In sufficiently large samples, ${k \choose 2}$ will be large enough that the coalescence rate due to genetic drift will be appreciable, at least until the number of lineages surviving back in time declines. 
From a technical standpoint, this is related to the question of whether the coalescent process ``comes down from infinity'' \citep[for a review see][]{Berestycki2009}. 

% Intuitively this occurs because each sweep is likely to affect at most two lineages, and so we are converging to a bifurcating tree similar to the neutral coalescent. 
\subsubsection*{The low frequency limit} \label{low_q_section}
As noted in our discussion of Figure \ref{Freq_spec_const_x_homogen}, the frequency spectrum may be close to neutral in appearance even with large reductions in $\pi$ if selected alleles sweep only to low frequency. In fact, by taking a limit (satisfying certain conditions)
in which sweeps occur frequently, but each sweep has a small probability of causing coalescence,
we can recover Kingman's coalescent.

We illustrate this limit by taking $\nu \to \infty$ and allowing $f(q)$ to depend on $\nu$ in such a way that as $\nu \to \infty$,
$I_{k,\ell}/I_{k,2} \to 0$ for all $3 \le \ell \le k$,
and that $\nu \; I_{k,2} \to {k \choose 2} \gamma$, for some $0 < \gamma < \infty$.
As shown in Appendix \ref{More_on_low_q}, a sufficient condition for this is that $\lim_{\nu \to\infty} \nu \, \int_0^1 q^2 f(q) dq$ is finite.
In this limiting case, the rate of coalescence is 
\begin{equation}
  \lambda_{k} = {k \choose 2} \left( \gamma + \frac{1}{2N} \right) \label{rate_low_q}       ,
\end{equation}
so the limiting model behaves exactly as the standard neutral coalescent but with an effective population size of
\begin{equation}
  N_e = \frac{2N}{2 N \gamma + 1 }. \label{Ne_sweeps} 
\end{equation}
Note that the limiting coalescent process does not satisfy condition \eqref{durrett_condition} of \citet{Durrett:05}, 
and that Tajima's $D$ and Fu and Li's $H$ will have mean equal to zero at all sample sizes,
as is natural since the limiting process is just the neutral (Kingman's) coalescent.

In the case of our simple partial sweep coalescent this limit would occur if the frequency $x$ reached by sweeps is taken to zero 
as the rate of sweeps grows at least as $1/x^2$.  The simple homogeneous full sweep coalescent process obviously can not be taken to this limit as there is a proscribed set of $J_{k,\bullet}$, which feature non-trivial amount of coalescence involving more than pairs of lineages. 
%% I put this back in as the reviewer 2 asks for the differences from the full sweep model

\paragraph{Interference}

In both limits discussed above the population-scaled rate of sweeps has to be very high. In the first limit the rate of sweeps has to be high enough to dominate the rate of neutral coalescence, in the second limit the rate of sweeps has to be high enough to compensate for the fact that any one sweep is very unlikely to cause coalescence. The requirement of a high rate of sweeps implies 
that interference between the sweeps may occur, thus violating our assumption that the sweeps are independent. 
Investigations of the effect of such interference on the signal of hitchhiking have shown that interference reduces the impact of any one sweep on patterns of polymorphism \citep{KimStephan:03,Chevin:08b},
although to interfere, the sweeps must begin at very similar times at loci separated by a low recombination rate.
This suggests that a very high rate of sweeps is needed indeed before interference will have an appreciable impact on the hitchhiking effect, 
as would occur in the homogeneous sweep model if $\nu_{BP}/r_{BP}$ is very large. 
The limits we describe above only require that the population size-scaled rate of sweeps ($N\nu$ or $N\nu_{BP}$) be high, 
and therefore it is possible to keep the {\em per generation} rate of sweeps sufficiently low as to avoid the effect of interference. Further work is needed to investigate coalescent models under such high rates of sweeps, 
and could be useful in understanding genealogical processes in organisms with low or no recombination that also experience strong selection pressures.

% GC agrees This seems unnecessary.
% Under parameter ranges where interference is likely to occur we may be able to account for its effects by modifying $J_{k,\cdot}$ to reflect the fact that some sweeps are prevented from achieving high frequency due to interference from other sweeps. However, to fully model the coalescent when interference is common we will need to model the temporal and spatial clustering of sweeps that interfere and the overdispersion of sweeps \plr{overdispersion of what?} that proceed unimpeded. 

\section{Discussion}
The prevailing view of adaptation in a population genetics setting is based on a lone selected allele racing from its introduction into the population to fixation, carrying with it a chunk of the chromosome on which it arose. This cartoon has been a very useful prop for developing tests to identify genes underlying recent adaptations, and for interpreting genome-wide patterns of polymorphism. However, it seems likely that such full sweeps constitute only a small proportion of the selected loci whose frequency changes in response to adaptation \citep[see][for a recent discussion]{Pritchard:10}. If we are to develop a better understanding of the full impact of linked selection on patterns of diversity we need to develop a richer and more flexible set of models.

The work in this paper was motivated by models in which the external environment or the genetic background vary on a fast enough time scale that new alleles rarely reach fixation before selective pressures change, either slowing their advance or reversing their trajectory. We laid out an approximation to the coalescent process under such a model, and showed that, while the initial rapid stage of the trajectory will strongly impact the coalescent process, subsequent slower dynamics of the selected alleles have a much smaller effect.
We then extended this idea to a recurrent sweep model, approximating the dynamics by a multiple-merger coalescent.
While some of our results are fairly general,
to provide a more intuitive sense we have often employed simple allele frequency trajectories and made other approximations.
Nonetheless, we expect more realistic models to give rise to qualitatively equivalent results.

Each sweep we consider consists of a single allele at a locus rising on a single haplotype from very low frequency in to the population. 
This contrasts with many other soft sweep models, 
under which a sweep starts on multiple haplotypes, 
either because multiple different alleles initially segregated at the locus \citep{Hermisson:05}; 
or as a result of multiple mutations occurring after selection pressures switched \citep{Pennings:06, Pennings:06b, Ralph:10}; 
or because the adaptive allele was previously neutral and present on multiple haplotypes \citep{Innan:04,Przeworski:05}. 
It is likely that recurrent models of such soft sweeps could be approximated through coalescent models with simultaneous multiple collisions \citep{Schweinsberg:00}, 
to model the simultaneous rise of multiple haplotypes.
This seems like a fruitful area of future work as it would substantially extend our understanding of the effects of a broad family of recurrent sweep models on genomic patterns of diversity.

%, due to chromosomes that carry deleterious alleles failing to contribute genetic material to the present day, 
We have also ignored the effect of background selection. 
To a first approximation, the effect of background selection can be modeled as an increase in coalescence rate,
which would be a minor modification to equations \eqref{k_to_i_w_drift} and \eqref{eqn:J_rates}.
This would alter the predicted relationship between diversity and recombination \citep{InnanStephan:03} given by equation \eqref{pi_I}, 
and would offer a simple way to model the genealogical effects of both general models of hitchhiking and background selection.

% While a number of the results provided apply to quite a general forms of selected allele trajectory, to provide a more intuitive sense of these results throughout this paper we have used a simple partial sweep trajectory. we emphasize that this simple trajectory is a prop to allow us to understand how the frequency achieved by sweep affects patterns of diversity.  More realistic models will lead to a richer set of dynamics in terms of the trajectories of selected alleles. However, given that the coalescent process at partially linked sites is relatively insensitive to many features of a trajectory,  these models will likely give rise to diversity patterns that are well described by the coalescent processes laid out here. 

\subsubsection*{The interpretation of population genomic patterns}

Models in which selective sweeps do not always sweep to fixation have a much wider spectrum of predictions than the recurrent full sweep model. 
Three broad correlations that have been used to argue for the prevalence of linked selection, 
and used to potentially discriminate between models invoking background selection or full sweeps are:
{\bf 1)} correlations between neutral diversity and the recombination rate;
{\bf 2)} correlations between the frequency spectrum and the rate of recombination; and
{\bf 3)} correlations between putatively adaptive divergence and neutral diversity. 
We now describe some of the implications of our results for understanding these patterns in population genomic data.

\paragraph{Correlation between recombination and diversity}
One of the earliest and most compelling pieces of evidence for the role of linked selection in the fate of neutral alleles is a positive correlation between recombination and levels of diversity at putatively neutral sites (factoring out substitution rates as a proxy for differences in mutation rate). This pattern is consistent with both full sweeps and background selection, as both predict positive, albeit differently shaped, relationships \citep{InnanStephan:03}. 
The shape of the diversity-recombination curve under a homogeneous rate of partial sweeps is identical to the full sweep model, and more generally for a broad class of homogeneous sweep models. 
In fact, the relationship under a homogeneous model only depends on $2N \nu_{BP}J_{2,2}$, as seen in equation \eqref{pi_I}.

To illustrate this point, in Table \ref{x_values} we present estimates of $2N \nu_{BP} J_{2,2}$ for humans and {\it Drosophila melanogaster}, 
assuming a model with drift and a homogeneous rate of selective sweeps across the genome,
and from equation \eqref{pi_I} and data from \citet{Hellmann:08,Shapiro:07}.
% The estimate for humans was taken  \citet{Hellmann:08}, who estimated an equivalent parameter under a full sweep model, 
% and the value for {\it Drosophila melanogaster} we estimated from the data of \citet{Shapiro:07}. 
Along with these estimates, Table \ref{x_values} also shows the implied rate of sweeps per generation per base pair, $\nu_{BP}$, 
under the simple partial sweep model, for a variety of values of $x$.
These rates are surely overestimates, are intended for illustrative purposes only, as they ignore the effect of other forms of linked selection, 
e.g.\ background selection. 

The strength of the relationship between diversity levels and recombination varies dramatically between the two species, 
as indicated by the very different estimates of  $2N \nu_{BP} J_{2,2}$ 
(note that the estimates of $\nu_{BP}$ are similar due to the thousand fold difference in $N$). 
In \textit{Drosophila} the positive relationship between recombination and diversity is strong \citep[e.g.][]{Aguade:89,Begun:92,Berry:91,Shapiro:07,Begun:07}, 
but in humans the relationship seems to be weaker and is and complicated by other confounding factors \citep{Payseur:02,Hellmann:03,Hellmann:05,Hellmann:08,Cai:09}. 
However, we should be cautious in the biological interpretation of this difference, 
as in humans diversity is usually estimated in large windows (much of which will be noncoding and far from genes), 
while in {\it Drosophila} neutral diversity levels are usually estimated from synonymous sites in individual genes. 
What is needed is a comparative analysis that studies these patterns at the same genomic scale 
and accounts for the profound differences in the density of functional targets among species. 

The fact that the diversity--recombination curve plateaus rapidly in humans 
is strong evidence that linked selection does not affect the average neutral site in regions of high recombination. 
Technically, this could also occur if the density of selective targets $\nu_{BP}$ decreases approximately linearly with recombination rate; 
however, this option does not seem likely {\it a priori}. 

Although in {\it Drosophila melanogaster} this curve is still concave, it does not appear to flatten completely in high recombination regions \citep[e.g.][]{Sella:09}, suggesting that linked selection is an important source of stochasticity even in these regions. At face value the concave nature of the curve suggests that both genetic drift and linked selection contribute to stochasticity, as $N\nu_{BP} \gg r_{BP}$ would lead to an almost linear relationship across the observed range of recombination rates (see equation \eqref{inf_hetero}). 
However, a model with effectively no genetic drift can produce a concave curve and fit the observed data if $\nu_{BP} J_{2,2}$ is not constant across recombination environments, e.g.\ if sweeps occur at a moderately higher rate or achieve higher frequency in high recombination regions.
Neither of these two options seem particularly unlikely, suggesting that we still have little unambiguous evidence favoring genetic drift as an important source of stochasticity in {\it Drosophila}.

\begin{table}

\begin{tabular}{|c|cc|cccc|}
\hline
& & & \multicolumn{4}{|c|}{$\nu_{BP}$ across a range of $x$} \\
 &$\theta$ & $2N\nu_{BP} J_{2,2}$ & $x=1.0$ & $x=0.5$ & $x=0.2$ & $x=0.05$ \\
\hline
%3.0e-12 1.2e-11 7.5e-11 1.2e-09
Human  &$0.0017$ & $ 6 \times 10^{-11}$ & $3.0 \times 10^{-12}$  &  $1.2\times 10^{-11}$ & $7.5\times 10^{-11}$ & $1.2\times 10^{-9}$\\
{\it D. mel} &$0.025$ & $ 7.3\times 10^{-9}$ & $3.6 \times 10^{-12}$ & $1.5\times 10^{-11}$ & $9.1\times 10^{-11}$ & $1.5\times 10^{-9}$\\
\hline
\end{tabular}

\caption{ 
{\bf Estimates of sweep parameters from the relationship between diversity and recombination.} The estimate for humans was taken from \citet{Hellmann:08} who fitted a curve of the form of equation \eqref{pi_I}. The estimate from {\it Drosophila melanogaster} ({\it D. mel}) was obtained from fitting equation \eqref{pi_I} to the synonymous polymorphism and sex-averaged recombination rates of \citet{Shapiro:07} (kindly provided by Peter Andolfatto, see \citet{Sella:09} for details) using non-linear least squares via the {\tt nls()} function in {\tt R}. These estimates were converted into estimates of the rate of sweeps per generation per base pair ($\nu_{BP}$, last four columns)  under the simple partial sweep trajectory model where $  J_{2,2} = x^2 / t_x$, assuming  $t_x =1,000$ generations (equivalent to a selection coefficient of $\sim 0.01$) and that $N=10^{6}$ in {\it D. mel} and $N=10^4$ in humans.}
 \label{x_values}
\end{table}
%, and a $\theta=2.75\%$ was estimated
% In the last two rows an estimate of $ 2 \mu_{BP} / (\nu_{BP} J_{2,2})$ under a sweep model without genetic drift is given for the {\it D. melanogaster} of data of \citet{Shapiro:07} partitioned into regions of high and low recombination rate and assuming  $\mu_{BP} \approx 1 \times 10^{-8}$ \citep{Haag-Liautard:07}  and the implied rate of partial sweeps is again given for a range of $x$.}

\paragraph{The frequency spectrum}
The recurrent full sweep model predicts a strong positive relationship between the reduction in neutral diversity and the skew towards rare alleles \citep{Braverman:95, Kim:06}, a pattern  not predicted under models of strong background selection. 
This relationship has been used to test between full sweeps and background selection models, 
although note that as we discussed in Section \ref{strong_limit}, this relationship is not expected if all coalescence comes from selective sweeps.
Under our simple trajectory model, the distortion of the frequency spectrum is primarily determined by the frequencies that sweeps achieve. 
Therefore, although a lack of a strong skew in the frequency spectrum is consistent with a low rate of full sweeps, 
it cannot be used to rule out a high rate of partial sweeps. 
A lack of a genomic relationship between the frequency spectrum and recombination rate is therefore not grounds for rejecting sweeps as a force in shaping genetic diversity in favor of a model of background selection. 
Our results suggest that recurrent partial sweeps to low frequency in regions of high recombination in {\it D. melanogaster} and in the low recombination regions in humans may be a major source of stochasticity in allele frequencies.

\paragraph{Correlation between divergence and polymorphism.} 
Attention has recently focused on examining the correlation between neutral diversity and amino acid substitutions (or other putatively functional changes) between recently separated species. If a reasonable fraction of amino acid substitutions are driven by new mutations sweeping to fixation, then levels of diversity should dip on average around amino-acid substitutions. This relationship has been tested for by looking for a positive correlation between diversity levels and amino-acid substitution rates \citep{Macpherson:07,Andolfatto:07,Cai:09,Haddrill:11} or for a dip in diversity levels around a large set of aggregated amino acid substitutions \citep{Hernandez:11, Sattath:11}. If the density of functional sites is properly controlled for, these types of correlations between amino-acid substitutions and neutral diversity are not expected under a (simple) model of background selection. Such correlations have been detected in \emph{Drosophila} \citep{Macpherson:07, Sattath:11} but in humans the dip in diversity around non-synonymous substitutions seems to result from the dip in diversity levels around genes, an observation that seems inconsistent with a high rate of strong full sweeps \citep{Hernandez:11}.  Similarly, it has been observed that the highest $F_{ST}$ signals between human populations are not associated with strongly reduced haplotypic diversity \citep{Coop:09}. 

The fact that selected alleles in the partial sweep coalescent model do not have to sweep all the way to fixation partially decouples the rate of fixation of adaptive alleles from their effects on patterns of diversity within populations.  Therefore, the strength of the positive relationship between substitution rates and diversity depends on the fate of alleles that sweep into the population. For example, this positive relationship may be weak, and a poor predictor of the total reduction in diversity, if the majority of adaptive alleles that initially sweep into the population are eventually lost  \citep[e.g.\ as can be the case for major effect alleles in polygenic models of adaptation, see][]{Lande:83,Chevin:08}. \\

\paragraph{Concluding thoughts} In this article, we have concerned ourselves with patterns of diversity at a single neutral site. However, partial sweeps also have a strong effect on linkage disequilibrium and haplotype diversity, a signature that has been exploited in scans for selection \citep[e.g.][]{Hudson:94, Sabeti:02,Voight:06}. One simple case that we can immediately describe is the low $q$ limit (section \ref{low_q_section}). In that limit, the coalescent is equivalent to the standard neutral model and as such the decay of LD will be the same as the standard neutral model with an $N_e$ given by equation (\ref{Ne_sweeps}). A natural way to extend this exploration would be the genealogical framework developed by \citet{Mcvean:07} that has recently been extended to a multiple mergers coalescent by \citet{Eldon:08}. 

We will soon have polymorphism data across a broad range of taxa that will differ dramatically in selection regimes, recombination rates, genome size, and population size allowing a much fuller picture of how these various factors interplay to shape genome-wide levels of polymorphism. The results presented here, however, suggest that we will continue to struggle to distinguish between modes of selection, as relaxing the assumptions of various models can generate a broad range of overlapping predictions. 

Despite that, our results suggest a promising way forward, since a broad range of sweep models can be captured by a simple parameterizations of multiple merger coalescence processes. Importantly, this would allow parameter inference under a general model of linked selection, rather than focusing on a limited number of specific models. For example, we could estimate the rate that selection forces different numbers of lineages to coalesce (parameterized by $\nu f(q)$) as function of recombination rates and the density of selective targets. As the multiple--mergers coalescent model is easily simulated under, it may be readily incorporated into many of our existing genealogical inference frameworks. It is likely that parameters of such models could be estimated very precisely from genome--wide data, allowing us to concentrate on what these high level summaries of polymorphism tell us about linked selection across genomic environments and species. Such inferences may be important if we wish to move beyond documenting the presence of linked selection towards describing the genealogical process in species where selection is a major source of stochasticity.

\subsubsection{Acknowledgements}
Thanks to Yaniv Brandvain, Chuck Langley, Molly Przeworski, Josh Schraiber, Alisa Sedghifar, and Guy Sella for helpful conversations and comments on previous drafts. We thank the two anonymous reviewers and the Editor for helpful feedback. This work is supported by a Sloan Fellowship and funds from UC Davis to GC and a NIH NRSA postdoctoral fellowship to PR.

\appendix
\section{Appendices}

\subsection{$J_{k,i}$ for a generalized trajectory} \label{I_generalized}

Recall that we defined in equation \eqref{homo_I}
\begin{equation}
  J_{k,i} = {k \choose i} \E_X \left[\int_0^{\infty} q(X,r)^i(1-q(X,r))^{k-i} dr \right],  ~~2 \leq i \leq k ,
\end{equation}
so that the rate at which the coalescent process having $k$ lineages coalesces down to $i$ lineages from ``selective'' events is $\nu_{BP}/r_{BP} J_{k,i}$.
The quantity $q(X,r)$ is the pathwise Laplace transform of the process $X$, defined in equation \eqref{general_q}, and consequently
\begin{equation}
1-q(X,r) = \int_0^{\infty} r e^{-r t}(1-X(t)) dt .
\end{equation}
It is useful to note that by changing the order of integration, 
\begin{align}
  \nonumber J_{k,i} &= 
  {k \choose i} \E_X \left[ \int_0^\infty \left( \int_0^\infty \cdots \int_0^\infty \prod_{j=1}^i X(t_j) \prod_{\ell=i+1}^k (1-X(t_\ell)) r^k \exp \left(-r\sum_{j=1}^k t_j \right) dt_1 \cdots dt_k \right) dr \right] \\
   &= k! {k \choose i} \E_X \left[\int_0^{\infty} \cdots \int_0^{\infty}  \frac{\prod_{j=1}^{i} X(t_j) \prod_{j=i+1}^{k} (1-X(t_j))}{\left(\sum_{j=1}^{n}t_i \right)^{k+1}} dt_1 \cdots dt_k \right]\label{appendix1}
\end{align}
for $2 \leq i \leq k$, as long as the integral is finite.
In the case of a pair of lineages $i=2$ and this simplifies to
\begin{equation}
J_{2,2} = 2 \E_X \left[\int_0^{\infty} \int_0^{\infty} \frac{X(\tau -t_1)X(\tau-t_2)}{(t_1+t_2)^3} dt_1 dt_2.\right] \label{appendix2}
\end{equation}

To briefly explore the conditions for $J$ to be finite, we will suppose that $X$ leaves zero as a power of $t$, 
i.e.\ $X(t) \sim t^\alpha$ for some $\alpha>0$, for small $t$. We see that $J_{k,2}$ increases as $\alpha$ increases,
i.e.\ the rate of sweeps is larger the more rapidly $X$ leaves zero.
In this case, $q(r) \sim C \, r^{-\alpha}$ for large $r$, where $C$ is a constant.
Then since
\begin{align*}
  J_{k,2} &= \lim_{L \to \infty} {k \choose 2} \int_0^L q(r)^2 (1-q(r))^{k-2} dr \\
    &\le \lim_{L \to \infty} {k \choose 2} \int_0^L q(r)^2 dr ,
\end{align*}
it can be seen that $J_{k,2}$ is {\em infinite} if $\alpha \le 1/2$, 
in the limit of an infinite genome.
More generally, if $X$ leaves zero more quickly than $\sqrt{t}$
(which may be biologically unrealistic),
then sweeps occurring arbitrarily far away along the genome will cause coalescences.

\subsection{Recursions to find $\E[T_{tot}]$ and $\E[T_1]$} \label{expected_times_append}

Two properties of interest are the expected total amount of time in the genealogy at a neutral locus ($\E[T_{tot}]$) and the expected total amount of time in terminal branches ($\E[T_{1}]$). 

We first derive the expected total time in the genealogy. Recall that if the coalescent process has $k$ lineages,
then it waits an exponentially distributed amount of time with mean $1/\lambda_k$,
and then jumps to a smaller number of lineages chosen with probabilities according to $p_{k,\ell}$,
with $\lambda_k$ and $p_{k,\ell}$ given in equations \eqref{tot_lambda} and \eqref{choose_event}.
Therefore, if we let $G_{n,k}$ be the probability that the coalescent process that starts from $n$ lineages ever visits the state with $k$ lineages, then
\begin{equation}
  \E[T_{tot}] = \sum_{k=2}^n \frac{k}{\lambda_k} G_{n,k}.   \label{num_tot} 
\end{equation}
By conditioning on the last state visited before dropping to $k$ lineages,
we can see that $G_{n,k}$ satisfies the recursion
\begin{equation}
  G_{n,k} = \sum_{i=k+1}^{n} G_{n,i} \; p_{i,k}, \quad \mbox{for} \;  k<n ,
\end{equation}
with $G_{n,n}= 1$.
This recursion is of upper triangular form, so is easily solvable,
which together with \eqref{num_tot} allows us to compute $\E[T_{tot}]$.

We now turn to the expected total time in terminal branches, 
i.e.\ those branches on which mutations will lead to singletons.
Note that, since all lineages are exchangeable,
$\E[T_1]$ is equal to $n$ times the mean time until a particular lineage -- say, the first one -- coalesces with any other.
To find this, 
let $S_{n,k}$ be the probability that at some point there are $k$ lineages,
and that one of those $k$ lineages is the original first lineage, still not coalesced with any others.
Then the mean time until the first lineage coalesces is $\sum_{k=2}^n \frac{1}{\lambda_k} S_{n,k}$,
and hence
\begin{equation}
\E[T_1] = n \; \sum_{k=2}^n \frac{1}{\lambda_k}    S_{n,k} \label{num_single} .
\end{equation}
As above, we can get a solvable recursion for $S_{n,k}$ by conditioning on the last coalescent event before reaching $k$ lineages.
If the coalescent process jumps from $\ell$ to $k$ lineages, then the probability that a given lineage is not part of this coalescent event is $(k-1)/\ell$,
and hence
\begin{equation}
  S_{n,k} = \sum_{\ell=k+1}^{n} S_{n,\ell} \, p_{\ell,k} \frac{k-1}{\ell}  \quad \mbox{for} \; k<n ,
\end{equation}
and $S_{n,n}=1$.
The recursion is also easily solvable, which lets us obtain $\E[T_1]$.

\subsection{More on the low $q$ limit} \label{More_on_low_q}

We would like to arrange things so that asymptotically, all coalescent events affect only two lineages.
We illustrate this limit by taking $\nu \to \infty$ and allowing $f(q)$ to depend on $\nu$ in such a way that as $\nu \to \infty$,
$I_{k,\ell}/I_{k,2} \to 0$ for all $3 \le \ell \le k$,
and that $\nu \; I_{k,2} \to {k \choose 2} \gamma$, for some $0 < \gamma < \infty$.
Since this model is a Lambda coalescent with $\Lambda(dq) = q^2 \nu f(q) dq + \delta_0(dq)/2N$,
if we rescale time by a factor of $C$,
a necessary and sufficient condition is that $ C \Lambda$ converges weakly to a point mass at 0.

To emphasize the dependence of $f$ on $\nu$ we write $f(q) = f_{\nu}(q)$ and $I_{k,\ell} = I_{k,\ell}(\nu)$.
We would like to find a simple condition under which the proportion of coalescences involving more than two lineages goes to zero,
i.e.\ that $I_{k,\ell}(\nu)/I_{k,2}(\nu) \to 0$ as $\nu \to \infty$ if $\ell > 2$.
Fix $k$, and suppose for convenience that $f(q)=0$ for all $q>1-\epsilon$, for some $\epsilon>0$.
Then 
\[
    \epsilon^k \int_0^1 q^\ell f_{\nu}(q) dq < \int_0^1 q^\ell (1-q)^{k-\ell} f_{\nu}(q) dq < \int_0^1 q^\ell f_{\nu}(q) dq ,
\]
so that $I_{k,\ell}(\nu)/I_{k,2}(\nu) \to 0$ if and only if
\[
  \frac{ \int_0^1 q^\ell f_{\nu}(q) dq }{ \int_0^1 q^2 f_{\nu}(q) dq } \to 0 .
\]
Using Jensen's inequality,
 \begin{align*}
   \frac{ \int_0^1 q^\ell f_{\nu}(q) dq }{ \int_0^1 q^2 f_{\nu}(q) dq } &\le \frac{ \left( \int_0^1 q^2 f_{\nu}(q) dq \right)^{\ell/2} }{ \int_0^1 q^2 f_{\nu}(q) dq } \\
     &= \left( \int_0^1 q^2 f_{\nu}(q) dq \right)^{(\ell-2)/2} ,
\end{align*}
so if $\int_0^1 q^2 f_{\nu}(q) dq \to 0$, this will be achieved.
By the same result,
\begin{align*}
  \frac{ I_{k,2}(\nu) }{ \nu {k \choose 2} \int_0^1 q^2 f_{\nu}(q) dq } \to 1,
\end{align*}
so that, rescaling time by a factor $C_{\nu}$,
if
\[
  \nu C_\nu \int_0^1 q^2 f_{\nu}(q) dq \to \gamma \quad \mbox{as} \; L \to \infty,
\]
then $\nu C_\nu I_{k,2} \to {k \choose 2} \gamma$ for all $k$.
In this limit, the rate at which a pair of lineages coalesces converges, and does not depend on the number of lineages present.

Ideally, we would illustrate this with an stochastic model for $X$.  
However, the formula requires the model to be analytically tractable to a degree satisfied by no population genetics models that we could think of,
and it is much easier to make a concrete choice of $f(q)$.
Consider the case where $f(q)$ is the density of a Beta($1,M$) distribution. 
The mean of this distribution is $1/(1+M)$.  In that case 
\begin{equation}
  I_{k,\ell} =  {k \choose \ell} \int_0^1 q^\ell(1-q)^{k-\ell+M-1} M dq = M {k \choose \ell} \bigg / {k+M-1 \choose \ell} ,
\end{equation}
so that as $M \to \infty$,
\[
M I_{k,2} = {k \choose 2}\frac{2 M^2}{(M+k-1)(M+k-2)} \xrightarrow{L \to \infty} 2 {k \choose 2},
\]
so if $\nu = M$, then $\gamma = 2$.
We can furthermore check that
\begin{equation}
  \frac{I_{k,\ell}}{I_{k,2}} = \frac{ {k \choose \ell} }{ {k \choose 2} } \frac{\ell!(k+M-\ell-1)!}{2!(k+M-3)!} \sim \frac{1}{M^{\ell-2}} \xrightarrow{M \rightarrow \infty} 0 .
\end{equation}
so that this simple case satisfies our limit.
\bibliography{partial_sweeps}
\end{document}